\documentclass[10pt]{iopart} 
\usepackage{iopams}
\usepackage{amsthm}
\usepackage{bbm}
\usepackage{psfrag}
\usepackage[dvips]{graphicx}
 
\newtheorem{example}{Example}
\newtheorem{theorem}{Theorem}
\newtheorem{sloppytheorem}{Sloppy theorem}

\newcommand\NN{{\mathbbm{N}}}

\newcommand\RR{{\mathbbm{R}}}
\newcommand\dd{{\mathrm{d}}}

\newcommand\smallo{{\mathrm{o}}}

\newcommand{\Heaviside}{\Theta}
\newcommand{\Dirac}{\delta}

\begin{document} 

\title{Energy landscapes and their relation to thermodynamic phase transitions} 

\author{Michael Kastner}
\address{$^1$ Physikalisches Institut, Universit\"at Bayreuth, 95440 Bayreuth, Germany}
\ead{Michael.Kastner@uni-bayreuth.de} 

\date{\today}
 
\begin{abstract}
In order to better understand the occurrence of phase transitions, we adopt an approach based on the study of energy landscapes: The relation between stationary points of the potential energy landscape of a classical many-particle system and the analyticity properties of its thermodynamic functions is studied for finite as well as infinite systems. For finite systems, each stationary point is found to cause a nonanalyticity in the microcanonical entropy, and the functional form of this nonanalytic term can be derived explicitly. With increasing system size, the order of the nonanalytic term grows unboundedly, leading to an increasing differentiability of the entropy. Therefore, in the thermodynamic limit, only asymptotically flat stationary points may cause a phase transition to take place. For several spin models, these results are illustrated by predicting the absence or presence of a phase transition from stationary points and their local curvatures in microscopic configuration space. These results establish a relationship between properties of energy landscapes and the occurrence of phase transitions. Such an approach appears particularly promising for the simultaneous study of dynamical and thermodynamical properties, as is of interest for example for protein folding or the glass transition.
\end{abstract}
%\pacs{05.70.Fh, 05.20.-y, 75.10.Hk} 


\section{Introduction}
\label{sec:intro}

Speaking about energy landscapes in physics, people typically refer to the study of stationary points (i.\,e., points of vanishing gradient) of an energy function (like the Hamiltonian or the potential) of a many-particle system. This concept has proved particularly useful for the investigation of {\em dynamical}\/ properties like reaction pathways in chemical physics or conformational changes in biophysics: having computed local minima of the energy landscape as well as transition states (i.\,e., stationary points with exactly one eigendirection of negative curvature connecting two minima), reaction pathways and reaction rates can be predicted from these data (see \cite{Wales} for a comprehensive textbook on the subject).

Since energy landscapes have been proved to be a useful concept in dynamics, it is tempting to try to understand also {\em statistical physical}\/ properties from the study of stationary points of a many-particle energy function: Thermodynamic equilibrium describes a physical system in the long-time limit, and therefore a link between dynamics and equilibrium statistical physics has to exist by definition. Understanding what this link looks like in detail has remained a challenge in theoretical and mathematical physics for decades, and the mathematical branch of ergodic theory has emerged from the efforts to tackle this problem. In this paper, we will not dwell on these issues, and will only take the mere existence of a link between dynamics and equilibrium statistical physics as a motivation: A concept that has been proved useful in dynamics might have a good chance of being useful in statistical physics as well. Therefore, we will investigate in this paper the effect of stationary points of a many-particle energy function on statistical physical properties, in particular on the existence or absence of phase transitions.

What is interesting and promising about this approach is that it will allow us to study both, statistical and dynamical properties from stationary points of an energy function. This should prove useful in particular for the study of physical phenomena like protein folding or the glass transitions where both, dynamical and statistical features play an important role.

\section{Microcanonical statistical physics}
\label{sec:microcan}

Consider a classical\footnote{For the moment the discussion is restricted to classical (i.\,e., non-quantum mechanical) systems. We will comment in the conclusions on how to extend the concepts to quantum systems.} Hamiltonian system of $N$ degrees of freedom, characterized by a Hamiltonian function $H(p,q)$ on phase space, where $p=(p_1,\dots,p_N)$ is the vector of momenta and $q=(q_1,\dots,q_N)$ is the vector of position coordinates. The most fundamental ensemble in equilibrium statistical physics is the microcanonical one, and it takes as a starting point the microcanonical entropy
\begin{equation}
s_N(\varepsilon)=\frac{1}{N}\ln\Omega_N(\varepsilon),
\end{equation}
where
\begin{equation}\label{eq:Omega_N}
\Omega_N(\varepsilon)=\int\dd p\dd q\,\Dirac(H(p,q)-N\varepsilon)
\end{equation}
is the density of states. The integration in \eref{eq:Omega_N} is over phase space, and $\varepsilon$ denotes the total energy per degree of freedom. In later sections we will look for signals of phase transitions in this microcanonical entropy. Since we have defined the entropy as a function of the energy, we will be able to observe only those transitions which are driven by the energy (or the temperature, which is the conjugate variable). To observe, say, a magnetic field-driven transition, we would need to consider the entropy as a function of the magnetization to which the field is conjugate, and all the results presented below had to be translated accordingly.

\section{Stationary points of $H$ and nonanalyticities of the entropy $s_N$}
\label{sec:finite}

A stationary point $(p_{\mathrm s},q_{\mathrm s})$ of the Hamiltonian function $H$ is defined as a point of vanishing gradient,\footnote{For such a gradient to exist, $p$ and $q$ are assumed to be continuous variables. This excluded discrete models like the classical Ising model from our discussion.}
\begin{equation}
\dd H(p_{\mathrm s},q_{\mathrm s})=0.
\end{equation}
The effect of a stationary point on statistical physical quantities is particularly pronounced in the microcanonical ensemble, as we will see in the following example.
\begin{example}\label{example:doublewell}
Consider the Hamiltonian function
\begin{equation}\label{eq:doublewell}
H(p,q)=\frac{1}{2}p^2+\frac{1}{4}q^4-\frac{1}{2}q^2,\qquad
p,q\in\RR.
\end{equation}
This double-well has two minima at $(p,q)=(0,\pm 1)$ with energy $\varepsilon=H(0,\pm1)=-1/4$ and a saddle point at $(p,q)=(0,0)$ with energy $\varepsilon=H(0,0)=0$ [see Fig.\ \ref{fig:doublewell} (left)]. The density of states $\Omega_1(\varepsilon)$ displays nonanalyticities precisely at these values of $\varepsilon$ [Fig.\ \ref{fig:doublewell} (right)].
\begin{figure}[htb]
\center
\psfrag{0.25}{}
\psfrag{0.75}{}
\psfrag{-0.25}{}
\psfrag{-0.75}{}
\psfrag{-}{}
\psfrag{Om}{\scriptsize $\Omega_1$}
\psfrag{u}{\scriptsize $\varepsilon$}
\psfrag{p}{\scriptsize $p$}
\psfrag{q}{\scriptsize $q$}
\parbox{4.5cm}{\includegraphics[width=4.6cm]{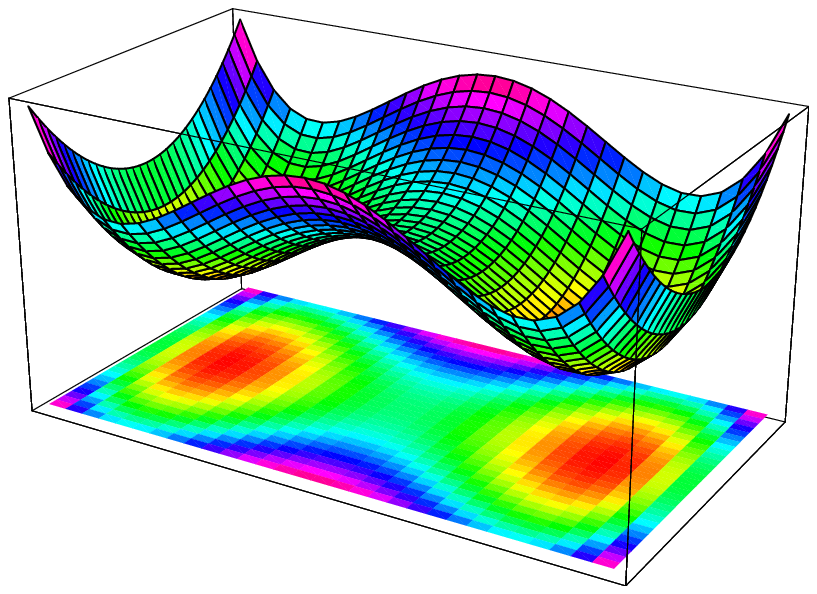}}
%\hfill
\parbox{3.8cm}{\includegraphics[width=3.8cm]{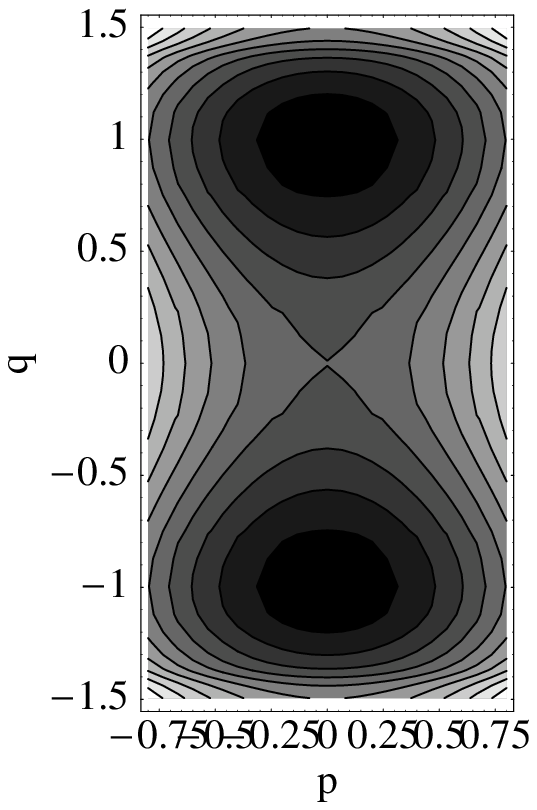}}
\hspace{-4mm}
\parbox{4.5cm}{\includegraphics[width=4.9cm]{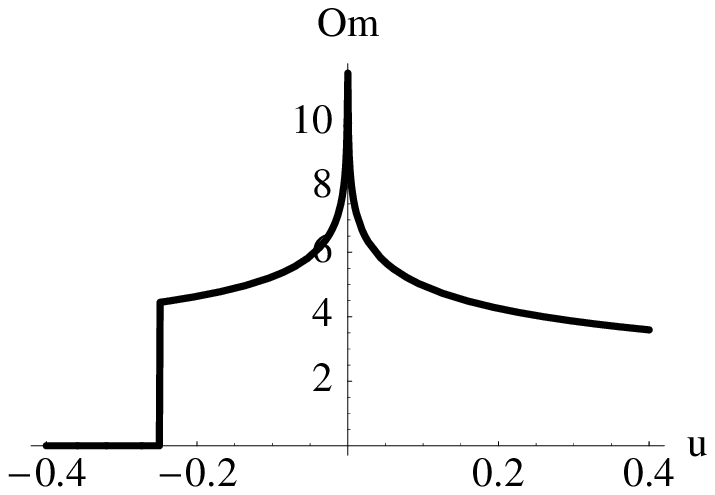}}
\caption{\label{fig:doublewell}
Plot of the Hamiltonian function (\ref{eq:doublewell}) as a three-dimensional plot (left) and as a contour plot (middle). The corresponding density of states $\Omega_1(\varepsilon)$ is nonanalytic at the energy $\varepsilon=-1/4$ of the minima and at the energy $\varepsilon=0$ of the saddle point of $H$ (right).}
\end{figure}
\end{example}
The relation between stationary points of the Hamiltonian function $H$ and the occurrence of nonanalyticities that we observe for this simple low-dimensional example turns out to hold true in arbitrary dimension and for stationary points of arbitrary index $i$.\footnote{The index $i$ of a stationary point $q_{\mathrm s}$ equals the number of negative eigenvalues of the Hessian ${\mathfrak H}_V$ of $V$ at $q_{\mathrm s}$.} To derive a quantitative result specifying this relation, it is convenient to make the following assumptions: Let $H$ be a Hamiltonian function of the standard form
\begin{equation}\label{eq:H_standard}
H(p,q)=\sum_{i=1}^N\frac{p_i^2}{2m_i}+V(q),
\end{equation}
where $m_i$ is the mass associated with the $i$th degree of freedom. The potential $V$ is a function of all position coordinates, and it may include external potentials, two-body interactions, or arbitrary $n$-body interactions. Since the kinetic energy term in (\ref{eq:H_standard}) is quadratic in the momenta, it is harmless as regards the occurrence of nonanalyticities in statistical physics.\footnote{Such quadratic terms give rise to Gaussian integrals in the partition function or density of states which can be solved, resulting in analytic contributions to the canonical free energy or microcanonical entropy.} Hence we can focus our attention in the following on the potential $V$ and study the configurational density of states
\begin{equation}\label{eq:Omega_conf}
\Omega_N(v)=\int \dd q\, \delta\left(V(q)-Nv\right)
\end{equation}
and the configurational microcanonical entropy
\begin{equation}\label{eq:s_conf}
s_N(v)=\frac{1}{N}\ln \Omega_N(v).
\end{equation}
We further impose the condition that $V$ be a so-called {\em Morse function}, requiring that, for all stationary points $q_{\mathrm s}$ of $V$, the Hessian ${\mathfrak H}_V$ of $V$ has a nonvanishing determinant,
\begin{equation}
\det {\mathfrak H}_V(q_{\mathrm s})\neq0.
\end{equation}
One may argue that this is an insignificant restriction, since Morse functions form an open dense subset of the space of smooth functions \cite{Demazure} and are therefore generic. This means that, if the potential $V$ of the Hamiltonian system we are interested in is not a Morse function, we can transform it into one by adding an arbitrarily small perturbation. One important consequence of the Morse property is that all stationary points of such a function are isolated which allows us to study the effect of a {\em single}\/ stationary point on the configurational density of states \eref{eq:Omega_conf}. Such an asymptotic analysis has been reported in \cite{KaSchneSchrei:07}:
\begin{theorem}\label{thm:finite}
Let $V:G\to\RR$ be a Morse function with a single stationary point $q_{\mathrm s}$ of index $i$ in an open region $G\subset\RR^N$. Without loss of generality, we assume $V(q_{\mathrm s})=0$. Then there exists a polynomial $P$ of degree less than $N/2$ such that at $v=0$ the configurational density of states (\ref{eq:Omega_conf}) can be written in the form
\begin{equation}\label{eq:Omega_sep}
\Omega_N(v)=P(v)+\frac{h_{N,i}(v)}{\sqrt{\left|\det\left[{\mathfrak H}_{V}(q_{\mathrm s})\right]\right|}}+\smallo(v^{N/2-\epsilon})
\end{equation}
for any $\epsilon>0$. Here $\Heaviside$ is the Heaviside step function, $\smallo$ denotes Landau's little-o symbol for asymptotic negligibility, and
\begin{equation}\label{eq:h_Ni}
\fl h_{N,i}(v)=\cases{
(-1)^{i/2} \,v^{(N-2)/2} \Heaviside(v) & \mbox{for $i$ even,}\\
(-1)^{(i+1)/2} \,v^{(N-2)/2}\,\pi^{-1}\ln|v| & \mbox{for $N$ even, $i$ odd,}\\
(-1)^{(N-i)/2} (-v)^{(N-2)/2} \Heaviside(-v) & \mbox{for $N,i$ odd.}\\
}
\end{equation}
\end{theorem}
For a proof of this theorem, the density of states is calculated separately below and above $v=0$. By complex continuation it is possible to subtract both contributions and to evaluate the leading order of the difference. A detailed proof of an even stronger result (including higher order terms) is given in \cite{KaSchneSchrei:08}.

The content of Theorem \ref{thm:finite} can be summarized as follows:
\begin{enumerate}
\item Every stationary point $q_{\mathrm s}$ of $V$ gives rise to a nonanalyticity of the configurational entropy $s_N(v)$ at the corresponding stationary value $v=v_{\mathrm s}=V(q_{\mathrm s})/N$.
\item The order of this nonanalyticity is $\lfloor (N-3)/2 \rfloor$, i.\,e., $s_N(v)$ is differentiable precisely $\lfloor (N-3)/2 \rfloor$-times at $v=v_{\mathrm s}$.
\end{enumerate}
It is interesting to note that the functional form of the nonanalytic term \eref{eq:h_Ni} does not depend on the precise value of the index $i$ of the stationary point, and the effect of a stationary point on the entropy is more or less independent of the index. Remarkably, in spite of this independence, several approaches have been reported in the literature in which statistical properties of many-particle systems are estimated exclusively from the minima of the energy landscape, disregarding all stationary points of higher index \cite{StillWe:82,Wales:93}. It is unclear to the author how these two observations fit together.

Numerical studies as well as heuristic arguments indicate that the number of stationary points of a generic potential $V$ increases {\em exponentially}\/ with the number of degrees of freedom $N$ \cite{DoWa:02}. As a consequence, for large (but finite) $N$ we can expect to find a very large number of nonanalyticities of $s_N$. Such behaviour is at pronounced variance to the properties of {\em canonical}\/ thermodynamic functions of finite systems which are known to be always analytic.

\section{Phase transitions in finite systems?}

In agreement with the intuition produced in Example \ref{example:doublewell}, we found a relation between stationary points of the many-particle potential $V$ and nonanalyticities of the configurational entropy $s_N$. But what is the physical significance of these nonanalyticities?

The microcanonical entropy is a so-called {\em thermodynamic function}, and the role that this function plays for the microcanonical ensemble is equivalent to that of the canonical free energy $f_N$ for the canonical ensemble. The canonical free energy is known to be an analytic function for all finite numbers $N$ of degrees of freedom \cite{Griffiths}, and it can develop a nonanalyticity only in the thermodynamic limit $N\to\infty$. These nonanalyticities of $f_N$ are of utmost importance in statistical physics and thermodynamics, and their presence is often taken as the defining property of a phase transition. But does it make sense to consider nonanalyticities of the microcanonical entropy as phase transitions of finite systems in the microcanonical ensemble?

I don't think so, the reason being that these nonanalyticities of $s_N$ do by no means show the remarkable properties which account for the interest of physicists in nonanalyticities of the canonical free energy. For typical many-particle systems, the nonanalyticities of $f_N$ (if they exist at all) are of very low order, meaning that already one of the first few derivatives is discontinuous. Such a nonanalyticity corresponds to a dramatic change of the physical properties due to collective effects. In contrast, the nonanalyticities of the finite-system configurational entropy typically occur at a huge number of different energy values, and their order decreases as the number of degrees of freedom becomes larger. The effect on the system's physical properties in that case is minor, and this is simply not the dramatic physical phenomenon which we would like to call a phase transition!\footnote{Note that here it was not our intention to define the finite-system analogue of a phase transition, but attempts of that kind have been published in the literature. The most widely used definition relates a first-order transition in a finite system to a convex intruder in the (otherwise concave) microcanonical entropy function, which is equivalent to an S-bend of the microcanonical caloric curve \cite{WaBe:94,Gross}.}

%Although we didn't even set out with the intention to define the finite-system analogue of a phase transition, it may be worth commenting on attempts of that kind in the literature. The most widely used definition relates a first-order transition in a finite system to a convex intruder in the (otherwise concave) microcanonical entropy function, which is equivalent to an S-bend of the microcanonical caloric curve \cite{WaBe:94,Gross}. Although such an approach can be useful in practice, it brings about several conceptual problems: First, upon variation of the energy, everything can remain smooth and well-behaved, which is not quite what you would expect for a phase transition. Second, similar to the weak nonanalyticities described in Sec.\ \ref{sec:finite}, the convex intruder in the entropy might become arbitrarily small with increasing $N$. Or there might be a convex intruder for all finite $N$, but a second-order transition in the thermodynamic limit. Third, you may find a convex intruder even for a system of, say, two degrees of freedom (maybe even one?), which would make it an effect which is not really a collective one. 

\section{Stationary points of $V$ and phase transitions in the large-$N$ limit}
\label{sec:tdl}

Having observed that stationary points of $V$ can be numerous and that the corresponding nonanalyticities of the microcanonical configurational entropy become weaker with increasing $N$, it may appear doubtful whether they are related to the occurrence of phase transitions in the thermodynamic limit at all. However, for several models, analytic calculations of all stationary points of $V$ have been performed, and the results give strong indications that a relation between these stationary points and phase transitions in the thermodynamic limit {\em does}\/ exist.
\begin{example}\label{ex:mf-ktrig}
One of those models for which all stationary points of $V$ can be computed analytically is the mean-field $k$-trigonometric model, characterized by the potential
\begin{equation}\label{eq:V_k}
V_k(q)=\frac{\Delta}{N^{k-1}}\sum_{i_1,\dots,i_k=1}^N \left[1-\cos\left(q_{i_1}+\cdots+q_{i_k}\right)\right],
\end{equation}
where $\Delta>0$ is a coupling constant and the coordinates $q_i\in[0,2\pi)$ are angular variables. The potential describes a $k$-body interaction where $k\in\NN$, and the model is known to undergo a phase transition for $k\geqslant2$ at a transition potential energy $v=\Delta$, whereas no phase transition takes place for $k=1$. In \cite{Angelani_etal:03} all stationary points of $V$ as well as the corresponding indices have been calculated. It is convenient to arrange these data in the form
\begin{equation}\label{eq:sigma}
\sigma(v)=\lim_{N\to\infty}\frac{1}{N}\ln\bigg|\sum_{j=0}^N(-1)^j\mu_j(v)\bigg|,
\end{equation}
where $\mu_j(v)$ is the number of stationary points $q_{\mathrm s}$ of $V$ with index $j$ and with a potential energy $V(q_{\mathrm s})/N$ not exceeding $v$. A plot of $\sigma$ is shown in Fig.\ \ref{fig:ktrig} (left).
\begin{figure}[ht]
\center
\psfrag{0.0}{\tiny $\!\!0.0$}
\psfrag{0.1}{\tiny $\!\!0.1$}
\psfrag{0.2}{\tiny $\!\!0.2$}
\psfrag{0.3}{\tiny $\!\!0.3$}
\psfrag{0.4}{\tiny $\!\!0.4$}
\psfrag{0.5}{\tiny $\!\!0.5$}
\psfrag{0.6}{\tiny $\!\!0.6$}
\psfrag{0.7}{\tiny $\!\!0.7$}
\psfrag{1.0}{\tiny $1.0$}
\psfrag{1.5}{\tiny $1.5$}
\psfrag{2.0}{\tiny $2.0$}
\psfrag{v/D}{\scriptsize $v/\Delta$}
\psfrag{s}{\scriptsize $\sigma$}
\psfrag{k=1}{\tiny $\!\!\!\!\!k\!=\!1$}
\psfrag{k=2}{\tiny $\!\!\!\!\!k\!=\!2$}
\psfrag{k=3}{\tiny $\!\!\!\!\!k\!=\!3$}
\psfrag{k=4}{\tiny $\!\!\!\!\!k\!=\!4$}
\parbox{6cm}{\includegraphics[width=6cm]{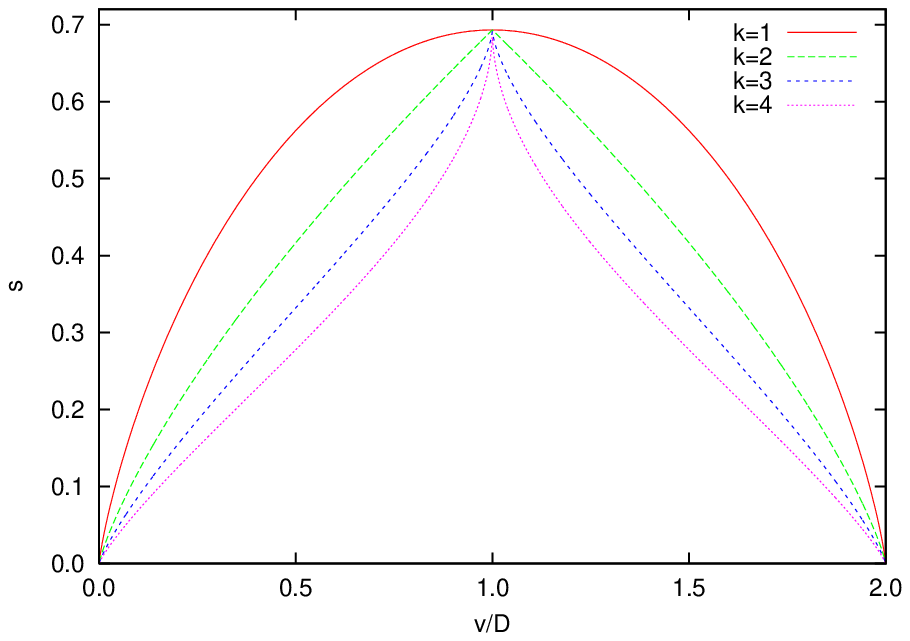}}
\hfill
\parbox{6cm}{
\psfrag{v}{\scriptsize $\!\!\!v/\Delta$}
\psfrag{j}{\scriptsize$ j_\ell$}
\psfrag{l}{}
\psfrag{-0.5}{\tiny $\!\!\!\!\!-0.5$}
\psfrag{ 0.0}{\tiny $\!\!0.0$}
\psfrag{ 0.5}{\tiny $\!\!0.5$}
\psfrag{ 1.0}{\tiny $\!\!1.0$}
\psfrag{ 1.5}{\tiny $\!\!1.5$}
\psfrag{ 2.0}{\tiny $\!\!2.0$}
\includegraphics[width=6cm]{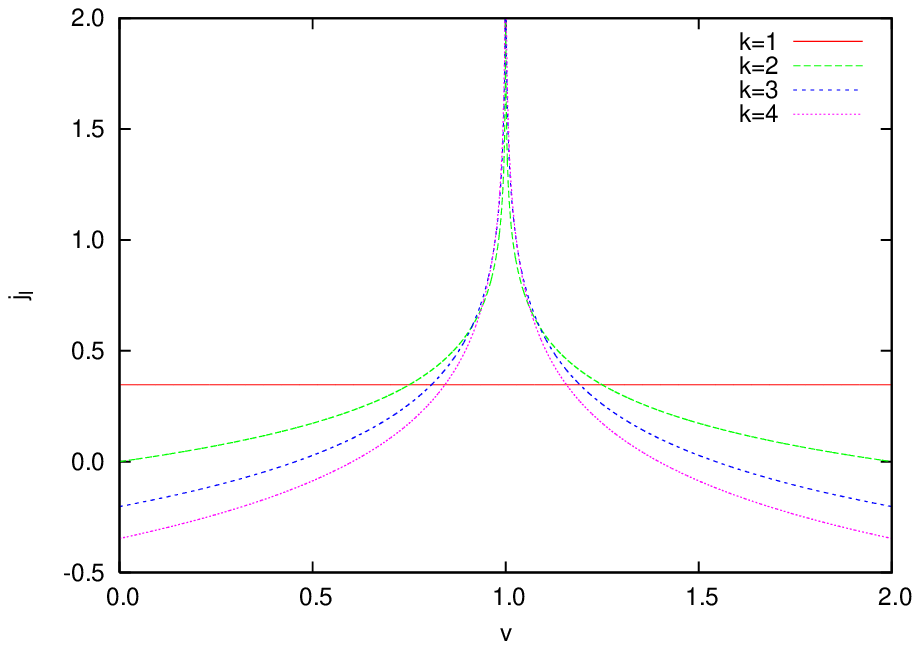}
} \caption{\label{fig:ktrig} Plot of the graph of $\sigma$ as defined in Eq.\ \eref{eq:sigma} (left) and of the `flatness indicator' $j_\ell$ (right) as functions of $v/\Delta$ for the mean-field $k$-trigonometric model in the thermodynamic limit. See Example\ \ref{ex:mf-ktrig2} for further details on $j_\ell$.}
\end{figure}
This quantity, containing exclusively information about the stationary points of $V$, already signals the absence or presence of a phase transition, being smooth for $k=1$ in the absence of a phase transition, and nonanalytic precisely at the phase transition energy $v=\Delta$ when $k\geqslant2$.
\end{example}
Other models for which the behaviour of $\sigma$ already indicates the presence of a phase transition are listed in Table 1 of \cite{Kastner:08}. Further evidence of a relation between stationary points of $V$ and the occurrence of phase transitions in the thermodynamic limit is provided by a theorem due to Franzosi and Pettini \cite{FraPe:04}. Although this theorem is originally stated in the language of topology changes of configuration space subsets, it can be rephrased in terms of stationary points of $v$. Here we give only a sloppy reformulation of this result:
\begin{sloppytheorem}\label{thm:FraPe}
Let $V$ be the potential of a system with $N$ degrees of freedom and short-range interactions. If some interval $[a,b]$ of potential energies per degree of freedom remains, for any large enough $N$, free of stationary values of $V$, then the configurational entropy $s(v)$ does not show a phase transition in this interval.
\end{sloppytheorem}
Note that a precise formulation of this theorem requires further technical conditions on the potential $V$ (see \cite{FraPe:04} for details).

Although stationary points of $V$ are necessary for a phase transition to occur, their presence is by no means sufficient. This is evident from Example \ref{ex:mf-ktrig} where stationary points were found to lie densely in an interval of the potential energy (per degree of freedom) axis, but a phase transition takes place at the single value of $v=\Delta$ only. To better understand the microscopic origin of a phase transition, we might now like to ask: {\em Which of the stationary points of $V$ may give rise to a nonanalyticity of the configurational entropy $s(v)$ in the thermodynamic limit?}\/ An answer to this question was given in Ref.\ \cite{KaSchne:08}. The idea behind this result is to sum up the nonanalytic contributions to the configurational entropy as given in Eqs.\ (\ref{eq:Omega_sep}) and (\ref{eq:h_Ni}) for all stationary points and to perform the thermodynamic limit of this sum. Then a bound on the magnitude of this sum is derived which can be interpreted in the following way:
\begin{sloppytheorem}\label{thm:flat}
The sum of the nonanalytic contributions of the stationary points of $V$ to the configurational entropy cannot induce a phase transition at the potential energy per particle $v=v_{\mathrm t}$ if, in a neighborhood of $v_{\mathrm t}$,
\begin{enumerate}
\item the number of critical points is bounded by $\exp(CN)$ for some $C>0$ and
\item the stationary points do not become `asymptotically flat' in the thermodynamic limit.
\end{enumerate}
\end{sloppytheorem}
`Asymptotically flat' here means that the determinant of the Hessian of $V$ at the stationary points goes to zero in the thermodynamic limit in some suitable sense (see \cite{KaSchne:08} for details). This result qualifies a subset of all stationary points of $V$ as `harmless' as what regards phase transitions and leaves only the (we hope few!) asymptotically flat ones as candidates for being at the origin of a phase transition. To illustrate the power of this theorem, we reconsider the mean-field $k$-trigonometric model from Example \ref{ex:mf-ktrig}.
\begin{example}\label{ex:mf-ktrig2}
Without going into the details, we will consider a function $j_\ell(v)$ which can be computed from the stationary points of $V$ and has the property of being divergent whenever stationary points with stationary value $v$ become asymptotically flat in the thermodynamic limit, and finite otherwise (see \cite{KaSchneSchrei:08} for details). For the mean-field $k$-trigonometric model (\ref{eq:V_k}), a plot of this function is shown in Fig.\ \ref{fig:ktrig} (right). In agreement with Sloppy theorem \ref{thm:flat}, $j_\ell$ is finite for $k=1$ where no phase transition takes place, but shows a divergence at the transition potential energy $v=\Delta$ for $k\geqslant2$. Despite the superficial similarity of the graphs of $j_\ell$ and $\sigma$ in Fig.\ \ref{fig:ktrig}, it is worth pointing out that, by virtue of Sloppy theorem \ref{thm:flat}, $j_\ell$ has a predictive power as what regards the occurrence of phase transitions which $\sigma$ is lacking.
\end{example}
With the results of the present section we have established conditions on the {\em microscopic level}, i.\,e., local properties in configuration space, which are relevant for the occurrence of a phase transition on the {\em macroscopic level}. Such conditions we may regard as a way of better understanding the origin of a phase transition.

Note that throughout Secs.\ \ref{sec:finite}--\ref{sec:tdl} we have made reference not to the density of states or the entropy, but to the configurational density of states or the configurational entropy as defined in \eref{eq:Omega_conf} and \eref{eq:s_conf}. The density of states is related to its configurational counterpart by means of a convolution (see Eq.\ (4.3) of Ruelle's textbook \cite{Ruelle}), and this relation allows us to translate our results on the configurational entropy into results on the entropy. For finite systems, we found in Theorem \ref{thm:finite} that the configurational entropy will typically display nonanalyticities of order $\lfloor (N-3)/2\rfloor$. As a consequence of the above mentioned convolution, nonanalyticities of the finite-system entropy will therefore be of order $N-1$, still increasing linearly with the system size $N$ \cite{CaKaNe}.

For a Hamiltonian function $H$ of standard form \eref{eq:H_standard}, every stationary point $q_{\mathrm s}$ of $V$ is in one-to-one correspondence to a stationary point $(0,q_{\mathrm s})$ of $H$. As a consequence, in the thermodynamic limit our sloppy theorem \ref{thm:flat} can be taken over verbatim to the entropy: The sum of the nonanalytic contributions of the stationary points of $H$ to the configurational entropy cannot induce a phase transition at the energy per particle $\varepsilon=\varepsilon_{\mathrm t}$ if, in a neighborhood of $\varepsilon_{\mathrm t}$, the number of critical points is bounded by $\exp(CN)$ for some $C>0$ and the stationary points do not become `asymptotically flat' in the thermodynamic limit.

\section{Conclusions and outlook}
\label{sec:conclusionsandoutlook}

\subsection{Conclusions}

In Secs.\ \ref{sec:finite} and \ref{sec:tdl}, we have studied nonanalyticities of the microcanonical entropy of finite and infinite systems, and in particular their relation to stationary points of the potential $V$. In summary we have seen that, for systems with standard Hamiltonian functions \eref{eq:H_standard}, the following statements hold true.
\begin{enumerate}
\item Stationary points of $V$ cause {\em nonanalyticities}\/ of order $\lfloor (N-3)/2\rfloor$ in the configurational entropy of {\em finite}\/ systems, at pronounced variance to the analytic behaviour of the canonical free energy density of finite systems.
\item The number of stationary points is believed to increase exponentially with $N$ for generic potentials $V$.
\item Due to the large number of finite-system nonanalyticities, their relation to phase transitions is clearly not one-to-one, but has to be somewhat more subtle.
\item In short-range systems, stationary points of $V$ are {\em necessary}\/ for a nonanalyticity to occur in the entropy, the configurational entropy, or the free energy in the thermodynamic limit.
\item Stationary points of {\em asymptotically vanishing curvature}\/ may cause a phase transition in the thermodynamic limit.
\end{enumerate}

\subsection{Quantum outlook}
\label{sec:outlook}

The concepts presented, based on stationary points of the potential energy function $V$, are of purely classical nature. However, a {\em geometric}\/ formulation of quantum mechanics (\cite{Kibble:79}; see \cite{AshSchil} for an introduction) provides a suitable framework to take over all the key features of these concepts to quantum mechanics. In this geometric framework, starting from a Hamiltonian operator $\hat{H}$ on Hilbert space ${\mathcal H}$, an energy expectation value function $h$ is defined on the quantum phase space ${\mathcal P}({\mathcal H})$, i.\,e., on the {\em complex projective space}\/ corresponding to ${\mathcal H}$. The function $h$ can be shown to have a number of stationary points which, remarkably, correspond to the eigenstates of the operator $\hat{H}$. Further elaboration and application of these concepts in a quantum context is currently in progress.

\subsection{Outlook on applications}

Up to now we have mostly emphasized the conceptual aspects of an analysis of stationary points of the potential energy $V$: We have gained insights into the analyticity properties of the microcanonical entropy and into the origin of phase transitions. Additionally, an analysis of the stationary points of $V$ is of interest in applications as well: For dynamical properties, as mentioned in the introduction, energy landscape methods have already been used extensively. Together with the link between stationary points and statistical physical properties presented in this article, it might be promising to investigate both, dynamical and statistical properties of a given system simultaneously on the basis of stationary points of the potential energy landscape. Such an approach should prove particularly useful for the study of protein folding or the glass transition where dynamical as well as statistical features play a role.

Both phenomena mentioned, the glass transition as well as protein folding, have already been investigated from an energy landscape perspective: For the glass transition, these studies were pioneered by Stillinger and Weber \cite{StillWe:82} who studied minima of the energy landscape, and generalizations to stationary points of higher index were reported by Grigera {\em et al.} \cite{Gri_etal:02} and others. Similarly, minima and other stationary points have been used for analyzing protein folding (see \cite{Wales} for an overview and a list of references). For both types of phenomena, the results suggest that studying stationary points of the energy landscape can be an efficient tool, but several difficulties have been encountered as well. The hope is that the results presented in this article will prove useful for identifying the properties or types of stationary points which are most relevant whenever entropic effects play a role.

\vspace{4mm}
\bibliographystyle{unsrt}
\bibliography{LandscapePT}

\end{document}